# Dynamical orbital evolution of asteroids and planetesimals across distinct chemical reservoirs due to accretion growth of planets in the early solar system.

Sandeep Sahijpal, Department of Physics, Panjab University, Chandigarh 160014, India
sandeep@pu.ac.in

**Abstract:**

N-body numerical simulations code for the orbital motion of asteroids/planetesimals within the asteroid belt under the gravitational influence of the sun and the accreting planets has been developed. The aim is to make qualitative, and to an extent a semi-quantitative argument, regarding the possible extent of radial mixing and homogenization of planetesimal reservoirs of the two observed distinct spectral types , viz., the S-type and C-type, across the heliocentric distances due to their dynamical orbital evolution, thereby, eventually leading to the possible accretion of asteroids having chemically diverse constituents. The spectral S-type and C-type asteroids are broadly considered as the parent bodies of the two observed major meteoritic dichotomy classes, namely, the non-carbonaceous (NC) and carbonaceous (CC) meteorites, respectively. The present analysis is performed to understand the evolution of the observed dichotomy and its implications due to the nebula and early planetary processes during the initial 10 Myrs (Million years). The homogenization across the two classes is studied in context to the accretion timescales of the planetesimals with respect to the half-life of the potent planetary heat source, $^{26}$Al. The accretion over a timescale of ~1.5 Myr. possibly resulted in the planetary-scale differentiation of planetesimals to produce CC and NC achondrites and iron meteorite parent bodies, whereas, the prolonged accretion over a timescale of 2-5 Myrs. resulted in the formation of CC and NC chondrites. Our simulation results indicate a significant role of the initial eccentricities and the masses of the accreting giant planets, specifically, Jupiter and Saturn, in triggering the eccentricity churning of the planetesimals across the radial distances. The rapid accretion of the giant planets, with appropriate eccentricities, critically influences the triggering of the orbital resonances that are in turn responsible for the radial mixing of the two distinct chemical reservoirs across the early solar system. This would influence the chemical composition and mixing of the various planetary reservoirs. The observed dichotomy among the NC and CC reservoirs can be preserved within the initial 5 Myr. in the early solar system in case the accretion of the two giant planets is prolonged. The present work provides a semi-quantitative formulation in terms of radial homogenization. A rigorous computational formulation of the evolving ensemble of distinct chemical reservoirs is beyond the scope of the present computational work.

1. Introduction

The present asteroid belt, at a heliocentric distance of ~2-4 AU (astronomical unit), consists of millions of survived rocky planetesimals that were formed in the early solar system. The gravitational collapse of the pre-solar molecular cloud ~4.56 billion years ago (Bouvier & Wadhwa 2010) led to the formation of the protosun at its center surrounded by an accretion disc of gas and dust. The physico-chemical processes associated with dust condensation, grain coagulation and gravitational accretion operating within the accretion disc during the initial few million years (Myr.) eventually led to the formation of planetesimals (Birnstiel *et al.* 2016). The majority of these planetesimals eventually either accreted to form the terrestrial planets and the cores of the giant planets (Scott 2007; Chiang & Youdin 2010; Kokubo & Ida 2012; Morbidelli *et al.* 2012; Johansen & Lambrechts 2017), or were dynamical ejected out of the solar system due to the gravitational perturbations of the planets that gradually enhanced their eccentricities, thereby, leading to their escape. The asteroid belt comprises of millions of such planetesimal remnants that survived over the eons against planetary accretion and dynamical ejection. The present work is an attempt to understand the dynamical orbital evolution of these bodies as planetesimals, specifically, during the initial 10 Myr. (million years) of the formation of the solar system in order to understand the origin of distinct populations of asteroids with different compositions. The chosen timescale corresponds to the duration during which the parent bodies of the meteorites were formed during the initial active solar nebula stage (Scott 2007; Dauphas & Schauble 2016), and experienced planetary scale thermal processing due to the potent radioactive heat source $^{26}$Al (e.g., Huss *et al.* 2006; Sahijpal *et al.* 2007; Alexander *et al.* 2018; Kleine *et al.* 2020; Morbidelli *et al.* 2022; Piralla *et al.* 2023).

Asteroids provide a unique gateway to understand the early solar system processes operating within the initial few millions years (Myr). Some of the earliest formed solar system pristine phases, e.g., Ca-Al-rich inclusions (CAIs), chondrules and metal condensates are found in chondrites that are derived from some asteroids. The elemental and isotopic analyses of these phases provide rare opportunity to understand the solar nebula and planetary processes operating during the initial few million years of the solar system that led to their formation and evolution. Among the asteroids, the spectral types, S-type and C-type asteroids are generally considered as the parent bodies of ordinary and carbonaceous chondrites, respectively (Morbidelli *et al.* 2015; Alexander et al. 2018). The lower abundance of organic and water contents in the former type compared to the latter suggests that the S-type asteroids were probably formed in the inner hot regions compared to the C-type asteroids forming in the outer cool regions of the solar nebula (see e.g., Morbidelli *et al.* 2015; Raymond & Izidoro 2017; Alexander *et al.* 2018; Morbidelli *et al.* 2022). The S-type asteroids are mostly confined to the inner asteroid-belt, whereas, the outer asteroid belt is dominated by C-type asteroids (Gradie & Tedesco 1982; DeMeo & Carry 2013, 2014; Morbidelli *et al.* 2015, 2016; Greenwood *et al.* 2020). Within their regions, the S-type and C-type asteroids have almost Gaussian shaped spatial distribution across the heliocentric distance, with a characteristic width of 0.5 AU (Morbidelli *et al.* 2015). The isotopic analyses of the bulk meteorites exhibit isotopic dichotomy between carbonaceous (CC) and non-carbonaceous (NC) chondrites and achondrites (Kleine *et al.* 2020; Schneider *et al.* 2020; Burkhardt *et al.* 2021) with well resolved distinct members. These two classes are broadly associated with the spectral types, C-type and S-type asteroids, respectively. Some of the recent works provide basic theoretical framework to understand the formation and evolution of elemental and isotopic dichotomy (Kleine *et al.* 2020; Schneider *et al.* 2020; Burkhardt *et al.* 2021) on the basis of the evolution of the two reservoirs within the solar nebula.

The chronological records of the early solar system phases, e.g., Ca-Al-rich inclusions, chondrules, bulk meteorites, etc., indicate that the active accretion phase of asteroids might have prolonged over the initial ~5 Myrs. (Scott 2007; Dauphas & Schauble 2016). However, some of the asteroids experienced rapid planetary scale melting and differentiation in the early solar system within the initial ~1.5 Myrs. due to the presence of the short-lived nuclide, $^{26}$Al (e.g., Sahijpal *et al.* 2007; Scott 2007; Kruijer *et al.* 2014, 2020; Kleine *et al.* 2020), that served as a potent heat source in the early solar system (Grimm & McSween 1989). The differentiated V-type and the M-type asteroids are considered to have provided a range of achondrites, iron and stony-iron meteorites, respectively. The V-type asteroids are referred as the Vestoids, with a spectral type identical to the asteroid 4 Vesta. These asteroids are known to be derived from the differentiated parent bodies (Hardersen *et al.* 2015; Morbidelli *et al.* 2016). The M-type asteroids are known to contain high metal contents on the basis of spectroscopy and are considered as the metallic cores of differentiated asteroids (Morbidelli *et al.* 2016; Shepard *et al.* 2015). The chondrite parent bodies evolved due to thermal metamorphism or aqueous alteration over longer timescales extending up to ~5 Myr and beyond (Kleine *et al.* 2020). The accretion of the planets also initiated during identical timescales.

The asteroid belt experienced dynamical evolution in terms of the orbital attributes, i.e., semi-major axis, eccentricity and inclination of asteroids since its formation. Apart from the accreting planets, the aggregate mass of the planetesimals within the asteroid belt in the early solar system influenced the dynamically evolution of the planetesimals (Wetherill & Stewart 1989; Wetherill 1992; Kokubo & Ida 2000, 2012; O'Brien *et al.* 2006, 2007), thereby, raising the eccentricities of the planetesimals by the massive objects accreted in the belt. Some of these associated works deals with the understanding of the dynamical evolution of planetesimals based on numerical simulations. The planetesimals smaller than ~10 km experience inward drift in gaseous disc on account of gas drag, whereas, the orbital motion of larger planetesimals remained uninfluenced by the gas drag (e.g., Capobianco *et al.* 2011). The acceleration experienced by a particle due to gas drag is inversely proportional to the size of the body. This makes smaller bodies more susceptible to drag. The drag coefficient experienced by small bodies could be two orders of magnitude higher than that experienced by larger bodies that have around unity drag coefficient due to their high inertia. Unless the small bodies are held together by localized turbulent currents or accreted to larger bodies, these bodies have a tendency to drift towards sun over a timescale much shorter than 1 Myr. The gravitational interaction of these small planetesimals probably resulted in the rapid growth of larger planetesimals that continues to gain mass by accreting more planetesimals. This is counter balanced by collisions that disrupt the growth. Several episodes of creation of the collision induced asteroid debris and re-agglomeration of the debris could have continued in a major manner at least over the life-time of ~10 Myrs. of the accretion disc during which the planets accreted bulk of their masses. The frequent collisions experienced by the accreting asteroids (diameter > 200 m) might have produced rubble-pile assemblages (Walsh 2018; Sahijpal 2022) as inferred on the basis of observed rotational velocities of the asteroids in comparison to the monolithic structure of smaller asteroids (Asphaug 2009). The asteroids in the main belt probably acquired a steady-state size distribution around a couple of billions years ago against collision (Bottke *et al.* 2015). The asteroids follow a power law distribution, with a power law index of ~3 (e.g., Asphaug 2009). The size distribution suggest that the asteroids with D (diameter) < 100 km represent collision induced disruption debris, whereas, the asteroid with D > 100 km are primordial that could have resulted from the accretion of smaller planetesimals. However, there is a possibility that the 100 km sized planetesimals were formed by direct gravitational collapse on account of streaming instability of turbulent pebble clouds rather than the accretion of small planetesimals (Klahr & Schreiber 2020). Further, a fraction of the asteroids with D < 30 km are lost from the belt due to Yarkovsky/YORP effects (Bottke *et al.* 2006).

There are broadly two distinct scenarios dealing with the origin and evolution of the asteroid-belt. The traditional hypothesis deals with the formation of planetesimals right within the belt, with the accumulated mass approximately 1000 times higher than the present mass (Wetherill 1980; Raymond *et al.* 2014). The mass depletion in the asteroid belt by three orders of magnitude (Morbidelli *et al.* 2009) cannot be explained exclusively by collision (Bottke *et al.* 2015). This mass depletion probably occurred very early in the history of the solar system along with the establishment of the eccentricities and inclination of the asteroids. The asteroids are presently associated with the eccentrics and inclinations in the range of 0-0.3 and 0-20º, respectively (Morbidelli *et al.* 2015). There are prominent Kirkwood gaps at $v_6$ secular resonance and 3:1, 5:2, 7:3 and 2:1 resonances as deciphered from observations and theoretically explained by numerical simulations (Minton & Malhotra 2009, 2011; Nesvorný *et al.* 2015). The majority of the resonances in the asteroid belt are due to the gravitational influence of Jupiter. In a simplified 3-body gravitational interaction regime with the sun, Jupiter and a specific asteroid/planetesimal, the asteroids/planetesimals with simple integer orbital period ratios with respect to Jupiter's orbital period experience substantial periodic orbital perturbation due to Jupiter. As a result these objects are gradually pushed into high eccentric orbits and are eventually lost from their initial orbits, thereby, resulting in the Kirkwood gaps. The gaps at 3:1, 5:2, 7:3 and 2:1 have orbital period resonance with Jupiter's orbit. The $v_6$ secular resonance is with respect to Saturn's orbit. There is a possibility that during the early stages of the gravitational interaction of the planetesimals with the planetary embryos the asteroid acquired the initial eccentricities and inclination (Wetherill 1992; Chambers & Wetherill 2001). This could have been later on modified by the gravitational interaction with the accreting planets.

The second hypothesis dealing with the origin of asteroids at wide-ranging heliocentric distances has gained more importance in the recent times (Raymond & Izidoro 2017; Raymond & Nesvorný 2021). The planetesimals/asteroids within this scenario were transferred from these regions to the present location in the asteroid-belt through the gravitational interaction of the accreting and migrating giant planets, Jupiter and Saturn. The two gaseous giant planets are considered to have substantially influenced the creation and dynamical evolution of the asteroid belt (e.g., Raymond & Nesvorný 2021). The formation of these planets started initially with the growth of 5-10 earth mass sized cores. These cores accreted gas during the gaseous accretion disc timescales of ~5 Myrs. (e.g., Piso & Youdin 2014). As one of the plausible scenario, it has been proposed that the C-type asteroids, along with Ceres, were gravitationally implanted into the outer asteroid belt from a widespread heliocentric distance of 4-9 AU by the accreting Jupiter and Saturn (Morbidelli *et al.* 2012). In this hypothesis, the compositional diversity of the carbonaceous chondrites can be explained to be originating from the widespread initial heliocentric distance of ~5 AU associated with the formation of planetesimals. The Grand Tack model that involves the inward migration of the accreting Jupiter and Saturn, followed by their outward migration, in the early solar system has been proposed as the major contributing factor for populating the asteroid belt (Walsh *et al.* 2011, 2012; Morbidelli *et al.* 2012; Raymond & Morbidelli 2014; Walsh & Levison 2016; Raymond & Izidoro 2017). The scenario is based on the strong gravitational influence of the migrating Jupiter and Saturn during their accretion stages in creating the present asteroid belt. In this scenario, the inward migration of accreting Jupiter and Saturn initially depletes almost the entire asteroid belt region of planetesimals due to their gravitational perturbation. This is followed by an outward migration of the two giant planets that have by now accreted substantial masses. The outward migration of the planets gravitationally populates mostly the inner asteroid belt with S-type asteroids from the inner region of the solar system that are also associated with terrestrial plant formation. As the giant planets migrate towards the outer solar system, their gravitational interaction populates the outer asteroid belt by the C-type asteroids/planetesimals that were initially present in the outer cooler regions of the solar system. The Grand Tack model provides a natural mechanism to populate the S-type and C-type asteroids preferentially in the inner and outer regions of the asteroid belt, respectively. Further, in comparison to the traditional view of the formation of planetesimals and planets around their present heliocentric distances (Wetherill 1980; Raymond *et al.* 2014), the Grand Tack model is able to explain the high Earth/Mars size ratio. It has been suggested that the Grand tack model leads to a rapid depletion of asteroids in the belt compared to the traditional scenario (e.g., Wetherill 1992; Morbidelli *et al.* 2015). The extent of homogenization among the diverse chondrites could have occurred during the active phase of population of the asteroid belt within the Grand Tack scenario and later-on during the accretion and migration of giant planets. Finally, the giant planet instability ~100 Myrs. after the formation of the solar system could have substantially changed the semi-major axis of the Jupiter, thereby, influencing gravitational interaction with the asteroids (Raymond & Nesvorný 2021).

In the present work, we have performed N-body numerical simulations of the orbital motion of planetesimals within the asteroid belt under the gravitational influence of accreting planets during the initial 10 Myrs. of the early solar system. The essential aim is to understand the dynamical orbital evolution of the asteroids and planetesimals due to the gravitational perturbation of accreting planets, thereby, resulting in the admixture of various asteroids spectral taxonomy classes found in distinct regions of the asteroid belt. These classes are broadly marked by C-type and S-type asteroids. The choice of the assumed timespan of the present simulation constitutes the primary assumption of the work (*Assumption #1*). We made this assumption since the meteoritic parent bodies were formed during this initial solar nebula stage, and experienced either thermal metamorphism, aqueous alteration, or melting and planetary scale differentiation due to the heat source $^{26}$Al present in the early solar system. Our primary objective is to understand how the planetesimals were dynamically evolved during this initial stage. However, this assumption completely ignores any further dynamical evolution of the asteroid belt in terms of the orbital motion and collisional history during the last 4.5 billion years (*Limitation #1*). In practice, it is possible to mathematically extend our simulations to any timespan but it is limited by our computational resources even with the use of parallel programing code. Thus, the present work provides a limited temporal window to the dynamical processes operating within the early solar system. However, this is precisely the relevant temporal window for this work.

We adopted the traditional scenario of the formation of asteroids within the asteroid belt (Wetherill 1980) with no contribution from the alternative scenario, the Grand Tack model (Raymond *et al.* 2014) (*Assumption #2*). This assumption imposes constraints on the initial conditions on the planetesimals of various compositions in terms of their initial heliocentric distances, specifically, in context with the Grand Tack model (*Limitation #2*). Our choice to simulate the traditional scenario is based on its numerical simplicity. However, we have tried to address the relevance and the consequences of the alternative model in the discussion section in a qualitatively manner. Following our traditional model, the planetesimals are initially introduced across the asteroid belt with assigned chemical composition among the two types, S-type and C-type. The orbits of these planetesimals eventually drift radially due to the gravitational perturbations of the accreting planets. On the basis of the detailed orbits an assessment at any given time can be made regarding the fraction of the each taxonomy classes to be present within any specific region. The collision induced fragmentation and re-accretion of these bodies would result in the ensemble evolution of these homogenizing compositions. Due to the anticipated intense computational complexities of these processes involving several thousands of bodies that are beyond the scope of the present work even though we have not performed the collisional induced chemical homogenization in a quantitative manner, yet we can present the estimates regarding the likelihood of the chemical homogenization due to hypothetical collision induced fragmentation and re-accretion of the evolving reservoirs with distinct compositions in a semi-quantitative manner (*Assumption #3*). The average chemical composition can be estimated based on the weighted average of the composition of the various planetesimals within a specific region that were gravitationally moved in from different initial heliocentric zones. This underlying assumption can be relaxed by considering a detailed computational study that will involve simulation of the collision induced fragmentation and re-accretion of several thousands of such bodies (*Limitation # 3*). This would be extremely computational extensive and has never been probably attempted. This remains one of our major limitations, and is elaborately discussed in the Methodology section. Based on our "semi-quantitative" numerical approach, the extent of compositional mixing of NC (S-type) and CC (C-type) rich planetesimals can be accessed as these planetesimals are energized by gravitational perturbations of accreting planets into high eccentric orbits, thereby, raising their collision probability with the planetesimals of distinct composition across the asteroid belt. Due to the possibility of the gradual collision induced accretion of asteroid populations from these mixing reservoirs, we can make assessment of the evolution of distinct asteroid chemical compositions during the initial 10 million years of the solar system. This could provide an insight, if not an actual account, into the chemical mixing evolutionary trends of the solar system due to possible accretion of planetesimals of distinct compositions. Since, $^{26}$Al is the main heat source for the thermal evolution of planetesimals/asteroids, it is important to understand the temporal evolution of the ensemble of these bodies along with the mixing of distinct chemical reservoirs. We performed the semi-quantitative compositional mixing analyses by taking into account, i) the timescale of the formation and evolution of the V-type and M-type asteroids (~1.5 Myrs.) that are considered to be the source of achondrites, iron and stony-iron meteorites, respectively, ii) the timescales for the thermal metamorphism and aqueous alteration in the asteroids (2-5 Myrs. or beyond), iii) the evolutionary timescales (>5 Myrs.) of the ensemble of rubble pile asteroids (Sahijpal, 2021, 2022, and references therein). During the initial ~1.5 Myrs. when $^{26}$Al was a potent heat source, the homogenization of the distinct chemical reservoirs at planetesimal scale could have produced homogenized differentiated parent bodies resulting in possible mixing across the two dichotomy NC and CC achondrites classes. The homogenization could have continued and produced parent bodies of meteoritic breccias with diverse fragments, clasts and mineral phases from different meteoritic classes extending from NC to CC chondrites. These bodies experienced thermal metamorphism or aqueous alteration during the initial 2-5 Myr. or beyond. The observed wide-range of meteoritic breccias indicates that the collision induced fragmentation and re-accretion of planetesimal debris of wide-ranging compositions and fragments was occurring in the solar system probably at a large scale (e.g., Bischoff et al., 2006). The present work provides a theoretical framework to understand the dynamical mixing of planetesimal debris of distinct compositions that could have continued even beyond the initial 5 Myrs, thereby, producing rubble-pile assemblages of various asteroids fragments. The various timescale presented above are based on the thermal modeling of meteoritic parent bodies and chronological records.

We have developed a parallel programming based numerical code to study the dynamical orbital evolution of the planetesimals/asteroids. The details are presented in section 2. The results are presented in section 3 along with the detailed discussion. The conclusions from the present work are finally presented in section 4.

## 2. Methodology

We have numerically simulated the dynamical orbital evolution of a population of asteroids of two distinct types, viz. the C-type and S-type by making several set of assumptions to make the simulations viable in terms of computational resources and speed. The asteroid chemical composition based on spectroscopy broadly indicate that the inner solar system, representing the asteroid belt, over time gradually acquired CC (C-type) composition at least in the outer asteroid belt, whereas, the inner belt acquired NC (S-type) composition. This assumption (*Assumption #4*) is supported by the observational spectroscopic records of the present asteroids along with the stable isotopic anomalies that indicate a compositional dichotomy in the early solar system (Kleine *et al.* 2020; Schneider *et al.* 2020; Burkhardt *et al.* 2021). In the absence of any viable observation records from the early solar system (*Limitation #4*) we cannot assume any alternative scenario to define the initial condition for our simulation. We adopted the traditional scenario of the *in situ* origin of planetesimals/asteroids within the asteroid belt (Wetherill 1980; Raymond *et al.* 2014) (*Assumption #2*). A defined number of planetesimals are initially introduced within the asteroid belt at the beginning of the simulations. The dynamical system of planets and planetesimals eventually evolves into an equilibrium state due to N-body gravitational interaction. A net depletion in terms of planetesimal numbers within the asteroid belt takes place on account of gravitational influence of the accreting planets. Due to the orbital resonance with Jupiter and Saturn, the planetesimals acquire high eccentricities. Eventually, some of these bodies are moved away from their initially assigned semi-major axis. In principle, some of these bodies can accrete on other planets. We have also estimated the probability of a body to cross the orbit of any planet, thereby, increasing its likelihood to be accreted by the planet. This is included in the tables provided in the supplementary file.

We divided the heliocentric region, representing the asteroid belt semi-major axis from 1.8 to 3.4 AU, into four concentric annular rings of 0.4 AU widths each. The assumption (*Assumption #5*) regarding the size of the four annular rings is based on the limitation of our computational speed. Further reduction in the ring widths will increase the simulation resolution but it will also substantially increase the computational time (*Limitation #5*). We tried to circumvent the limitation by running two set of simulations with varied configurations of the initial distribution of S-type and C-type asteroids. We do not have any direct observation based knowledge regarding the initial heliocentric distribution of the two asteroid types at the beginning stages of the formation of the asteroid belt except for the basic fact that the NC and CC bodies are presently concentrated in the inner and the outer regions, respectively. Hence, we considered two set of initial conditions (Table 1), along with a prescribed initial number of asteroids, for the heliocentric stratification within the traditional scenario. The underling understanding is that over time one of these initial sets of stratification could eventually dynamically and chemically evolve into the present observed heliocentric stratification (Morbidelli *et al.* 2015). In the set-I simulations (Table 1), the inner $1^{st}$ annular ring is assumed to initially have NC composition, whereas, the outer $2^{nd}$ - $4^{th}$ annular rings have an initial CC composition. In the set-II simulations, the inner $1^{st}$ and $2^{nd}$ annular rings are assumed to initially have NC composition, whereas, the outer $3^{rd}$ and $4^{th}$ annular rings have an initial CC composition. The two scenarios with distinct initial composition distributions would appropriately represent the end-members for the observed heliocentric distribution of asteroids of the two major spectral taxonomy classes. Within the traditional scenario, in the absence of any other proposed viable alternative, the two adopted initial distributions seem to be the most favorable options for the origin and evolution of asteroids within the asteroid belt.

The Table 1 also describes the nature of the simulations in terms of the state of the major planets. The simulation sets AI and AII deals with the contemporary state of the interaction of the accreted planets with the asteroids. The numerical code was essentially developed with these sets of conditions. These two simulations establish the robustness of our numerical code, and indicate the behavior of the system if the asteroids and fully accreted planets with their contemporary eccentricity values are introduced at the beginning of a simulation. These simulations were run with 2908 initial number of test particles (planetesimals) per ring for a time-span of 5 Myr. The asteroid belt establishes a steady state distribution within the initial couple of million years with significantly no observable subsequent change from 4-5 Myr. The robustness of the simulations was also confirmed by lowering the number of test particles by a factor of 5 in another simulation. The results of this simulation are not presented here. The Kirkwood gaps along with other features in the asteroid distribution were also observed in this simulation. On the basis of these simulations, we optimized the number of test particles per ring to 1088 in the remaining simulation sets BI, BII and CI, CII so that the simulation can be run for a longer time-span of 10 Myr. These two simulation sets constitute our major work as these simulations deals with the gravitational influence of the accreting planets on the planetesimals and asteroid in the early solar system. The eccentricities of the accreting planets were assumed to be zero in the set BI and BII, whereas, we adopted the present values of the eccentricities for the planets in the set CI and CII. The two sets cover the extreme possible range in the eccentricity values for the accreting planets.

Even though, during the nebula stage the processes related with condensation and thermal processing of Ca-Al-rich inclusions, chondrules, metal grain and matrix occur probably within the inner solar system, we have not numerically simulated any such processes as it is beyond the scope of the adopted numerical technique. The present simulation adopts the commencement of the accretion of planetesimals from the pebbles of distinct composition (Johansen & Lambrechts 2017) right from the beginning of the simulation and their subsequent dynamical orbital evolution. Pebble accretion scenario is generally considered as the most preferred mode for the accretion of planetesimals and planets. We have not performed any such simulation from pebble to planetesimal stage. Further, on the basis of the earlier work, it is assumed that the formation and/or the accretion of the simulation planetesimals could occur over a timescale of ~$2 \times 10^5$ years (Kokubo & Ida 2012). Each planetesimal introduced within the four annular rings is individually tracked for the entire simulation. These planetesimals are assigned a specific composition at the beginning of the simulation as discussed earlier (*Assumption #3*). Within this assumption, we cannot numerically simulate the collisions among planetesimals followed by fragmentation and re-accretion of the planetesimals. This is difficult to achieve numerically. The probability of a single collision ($P_{Coll.}$) among two distinct planetesimals initially originating from the $i^{th}$ and the $j^{th}$ annular rings, respectively, within a $k^{th}$ annular ring is proportional to the product of the probabilities, $P^i_k$ and $P^j_k$, of finding these planetesimals within the ring at a specific time (Equation 1).

$$P_{Coll.} \alpha\ P^i_k \times P^j_k \qquad (1)$$

Here, the indices, i, j and k, represent all the possible combinations of the four annular rings. Even though, the present work makes an assessment regarding the probabilities, $P^i_k$ and $P^j_k$, yet it is difficult to follow the dynamical collision courses. Further, we cannot numerically simulate the processes related with collision, fragmentation and re-accretion. In general, the fragments generated from several such collisions could re-accrete over time to produce planetesimals of varied compositions depending upon the weighted averages of the compositions of the contributing debris. In the absence of this numerical possibility what we can achieve is a hypothetical possibility that in case collisions, followed by fragmentation and re-accretion of the planetesimals were allowed to happen within a region, the system in equilibrium will achieve a homogenized chemical composition based on the distinct compositions of the initially introduced planetesimals within the region. On average, the re-accreted planetesimals could have achieved homogenized chemical composition. If $f^i_k$ represent the mass fraction of the collision debris contribution in the $k^{th}$ annular ring from the planetesimals that were initially introduced in the $i^{th}$ annular ring, and experienced an orbital drift, the normalized bulk weighted average composition $X^m_k$ for any isotope, 'm' of the re-accreted planetesimals in the $k^{th}$ annular ring from the contribution of debris from all the four annular rings, subsequent to the homogenization of the reservoirs, can be presented by Equation (2).

$$X^m_k = \Sigma^{i=1,4}\ f^i_k \times x^m_i \qquad (2)$$

Here, $x^m_i$ represent the normalized bulk composition of the same isotope for the planetesimals that were initially introduced in the $i^{th}$ annular ring and experienced an orbital drift. The indices, i and k, represent the four annular rings. The fractions $f^i_k$ will be proportional to the probabilities, $P^i_k$ that have been estimated numerically in the present work using the simulations. Thus, in a semi-quantitative manner, we can access the extent of homogenization among the planetesimals initially originating from distinct chemical reservoirs. Subsequent to the re-accretion of the debris from the planetesimals from these reservoirs, these fractions will survive in the form of various meteoritic phases in the multi-breccia. A possible melting of these re-accreted planetesimals by $^{26}$Al within the initial ~1.5 Myrs. could produce achondrites or iron-meteorite parent bodies with homogenized composition across the observed dichotomy (Kleine *et al.* 2020; Schneider *et al.* 2020; Burkhardt *et al.* 2021). The re-accretion of these bodies beyond ~1.5 Myr. would result in thermal metamorphism, aqueous alteration or rubble-pile formation of these multi-breccia bodies. Since, the system need not achieve complete chemical homogenization in terms of collision induced fragmentation and re-accretion among all the planetesimals present within a region at any instance, we can only associate the most likelihood probability for the planetesimals, in a semi-quantitative manner, at any given epoch of the simulation to acquire the prevailing chemical composition that they could have achieved in equilibrium. An epoch in the present work represents time from the beginning of the formation of the solar system, marked by the condensation of the earliest Ca-Al-rich inclusions in nebula, to a specific time during the initial 10 Myr. Since, the early solar system was experiencing a wide range of physical processes during the formation of the planetesimals, it is quite possible that the semi-quantitative formulation proposed above could be operating during the accretion, collision induced fragmentation and re-accretion stages of the planetesimals. Thus, the present simulations deal with two important aspects, viz., i) the numerical prediction of the orbital motion of a swarm of planetesimals from distinct initial heliocentric distances under the gravitational influence of sun and the accreting planets. ii) the likelihood of assigning an averaged chemical composition to a specific planetesimal at a time O(5) following a possible hypothetical homogenization on the basis of the average composition prevailing in the ring. However, the number of planetesimals introduced at the beginning of the simulations remains same during the simulation runs except with a wide-ranging spread in the semi-major axis and eccentricities. In the next section, we have also qualitatively discussed the possibility in case the system had not achieved the assumed equilibrium. .

The dynamical orbital evolution of the asteroids/planetesimals was performed by making several additional simplified assumptions in order to reduce the computational time and numerical complexities. The simulations were performed on a computational facility at IUCCA, Pune, in a parallel processing mode. All the planetesimals within the asteroid belt were considered to have a normalized test unit mass (1 unit) with an initial NC and CC compositions as described above (*Assumption #6*). The assumption of unity mass has also been adopted in most of the earlier works in the field. This assumption implies that the gravitational influence among the planetesimals was completely ignored and the planetesimals experienced only the gravitational influence of the sun and the planets. A detailed numerical analysis with a proper mass distribution and gravitational interaction among the planetesimals is simply not possible with the computational resources (*Limitation #6*). However, this will not drastically influence the major inferences drawn from the present work as majority of the planetesimals would contribute insignificantly to the gravitational interaction due to their low masses compared to sun and the massive planets, Jupiter and Saturn that essentially determines the dynamics of the asteroids even in the present solar system.

The planetesimals were introduced in their orbits in the beginning of the simulations and were confined to the elliptical plane, the Earth's orbital plane around the sun. We assumed a zero inclination *(Assumption #7)* for the orbits of asteroids as this will not substantially influence the major outcome of the present work since we assume 2-D broad annular rings of 0.4 AU width each for defining the distinct chemical reservoirs and the orbital crossing across these rings. We will have to eventually project the 3-D orbital motion on this 2-D plane even if we include inclination in order to do the homogeneity analysis. An inclination in the range of 0-20º has been observed for the present asteroids that can be associated with asteroid family clusters for lifetimes smaller than the age of the solar system (Asphaug 2009). The assumption of zero inclination was also made to increase the computational speed. This reduces the three-dimensional (3-D, X-Y-Z) dynamical problem to 2-D (X-Y) dynamical problem, thereby, improving the computational speed considerably. The incorporation of inclination would result in the dynamical evolution of the asteroid family clusters over the evolution of the solar system (*Limitation #7*). However, the planer nature of the annular rings in the present work does not necessitate the need.

The orbits of the planetesimals were determined based on the gravitationally influence of the sun and the eight accreting planets within the regime of Newtonian mechanics. It was assumed that the gravitational influence of the dwarf planets and the trans-Neptunian objects does not directly influence the orbital motion of asteroids *(Assumption #8)*. However, according to the Nice model (Gomes *et al.* 2005), the initial pool of the trans-Neptunian objects substantially influence Jupiter and Saturn, resulting in their outward migration (*Limitation #8*). Since, we have not included this aspect of giant planet migration in the present work hence, this will not influence the dynamics of planetesimals at least within the initial 10 Myr. This, however, constitutes one of the limitations that could be computationally difficult to address at the scale of solar system. This kind of study has probably never been carried out so far.

As mentioned in the introductory section, the gaseous accretion discs around the protostars generally last for ~5 Myrs. (Pfalzner *et al.* 2014). Thus, the gas drag on smaller bodies (< 10 km) becomes relevant at least for the initial 5 Myrs. The influence of gas drag was not considered in the present simulations *(Assumption #9)*. The gas drag will result in substantial migration of small bodies < 10 km towards sun, thereby, resulting in the rapid homogenization of CC bodies with the NC bodies unless this is circumvented by turbulent currents within the nebula disc that maintain local heterogeneities, or the smaller bodies rapidly accrete into larger bodies. This will introduce limitations in the orbital dynamics of small bodies that could play a major role in chemical homogenization (*Limitation # 9*). Further, the Yarkovsky effect is much more complicated to anticipate. It is responsible for the rapid drift of small bodies due to radiative heat losses into the Kirkwood gap from where these bodies are moved out rapidly due to the gravitational perturbation of giant planets. We have also not considered the dynamical depletion of the asteroids/planetesimals due to Yarkovsky effect *(Assumption #10)*. This would also delimit the dynamical role of small bodies in bringing chemical homogenization. However, we do not anticipate this effect in a substantial manner in the early solar system, specifically, during the initial 5 Myr. as the direct solar radiation is shielded by the solar nebula and does not substantially reach the planetesimals embedded within the nebular. The last two assumptions cannot be relaxed in the present work due to the extensive computational requirements along with the requirement to include additional physical processes that would drastically slow down the simulation. The addition of even a single equation associated with a specific process substantially increases the computational time due to the iterative nature of the simulation that has to be compromised by significantly reducing the number of test particles in a simulation. It should be, however, noted that majority of the earlier simulation works generally ignore these physical processes unless these works are specifically focused on these issues. A wide-range of numerical techniques has been adopted earlier in these works for numerically simulating the dynamical evolution of the solar system bodies, especially, the asteroids and planetesimals (e.g., Morbidelli 2002; Minton & Malhotra 2011; Walsh *et al.* 2011; Wisdom 2017). A N-body numerical code was developed in the present work on the basis of estimating the net gravitational force experienced by test particles (planetesimals) and planets. The corresponding acceleration experienced by the planetesimals due to sun and the planets were estimated at an integration time-step of 8 hours. This time-step was optimized during the development of the numerical code. The velocity and spatial displacement vectors were calculated on the basis of the acceleration vectors. The Equations (3-11) present the dynamical evolution of planets and planetesimals.

$$r_i \equiv r_i(r_{xi}, r_{yi}) \; \forall \, i > 10, \text{N planetesimals}$$

$$r_j \equiv r_j(r_{xj}, r_{yj}) \; \forall \, \text{sun (j = 1) \& the eight planets (j = 2, 9)} \tag{3}$$

$$v_i \equiv v_i(v_{xi}, v_{yi}) \tag{4}$$

$$a_i \equiv a_i(a_{xi}, a_{yi}) \tag{5}$$

$$a_{xi} = \sum_{j=1}^{9} G M_j (x_j - x_i) \left[ (x_j - x_i)^2 + (y_j - y_i)^2 \right]^{-3/2} \tag{6}$$

$$a_{yi} = \sum_{j=1}^{9} G M_j (y_j - y_i) \left[ (x_j - x_i)^2 + (y_j - y_i)^2 \right]^{-3/2} \tag{7}$$

$$v_{xi} = v_{xi} + t \, a_{xi} \tag{8}$$

$$v_{yi} = v_{yi} + t \, a_{yi} \tag{9}$$

$$r_{xi} = r_{xi} + t \, v_{xi} \tag{10}$$

$$r_{yi} = r_{yi} + t \, v_{yi} \tag{11}$$

The initial position vectors of the (N-10) number of planetesimals (Equation 3), assigned by an index, 'i', introduced within the four annular rings were randomly assigned according to the prescription mentioned below in the text. The planets were introduced with their masses, initial eccentricities and semi-major axes according to the adopted simulation criteria mentioned in Table 1 for the three set of simulations. A random set of values for the initial semi-major axis and eccentricities were chosen for the planetesimals as an initial condition. The values of the initial eccentricities were randomly selected in the range of 0-0.2 following a flat distribution of random numbers. The choice is based on the presently adopted Monte Carlo numerical approach that postulates that physical systems acquire equilibrium over evolution. It was assumed that the planetesimals had already acquired these eccentricities either through gravitational interaction or with the growth of planetary embryos (*Assumption #11*). This assumption has been generally adopted in the majority of the earlier simulation works where the observed eccentricity spreads of the present asteroids were adopted as the initial condition. An initially assumed zero eccentricity of asteroids fails to further excite the eccentricity during the dynamical evolution. The gravitational interactions among the planetesimals are generally considered to have created the initial spread (Wetherill 1980). The position vectors of the planets, assigned by an index, 'j' (Equation 3), were also defined according to their present semi-major axis. The semi-major axis of the planets Mercury to Neptune, assigned by j = 2, 9, in order of their heliocentric distances, were assumed to be 0.386, 0.724, 1, 1.52, 5.2, 9.58, 19.2, 30 AU (1 AU = $1.5 \times 10^{11}$ m), and the present eccentricity values of 0.048, 0.057, 0.047 and 0.00859 for Jupiter, Saturn, Uranus and Neptune, respectively, were assumed in the simulations according to the Table 1. The sun, assigned by j = 1, was assumed to be at the center of the co-ordinate system. The initial velocity vectors (Equations 4) of the planetesimals and planets were estimated using Kepler's law for the various planetary bodies. The acceleration vectors (Equation 5) were estimated using Equations (6-7) at every time-step, 't', of the simulations. This value is assumed to be 8 hours based on our experience gained during the simulation runs on the basis of the numerical consistency. The acceleration vectors for every planetesimal, with an index, 'i', are calculated by considering the contribution of the sun and all the planets. The G in the Equations (10-11) represents gravitational constant. In a similar manner, the acceleration vectors of all the planets were estimated by considering the contribution of the sun and the remaining planets. The velocity vectors of the planets and planetesimals were modified based on the estimated acceleration vectors using the Equations (8-9). Based on these velocity vectors, the new positions vectors were estimated for all the bodies using the Equations (10-12). The gravitational acceleration vectors were re-estimated for the new position vectors. The dynamical evolution governed by the Equations (6-11) was deduced in an iterative manner to simulate the orbits of the planets and planetesimals. The orbital motions of these bodies were analyzed to estimate the changes in their semi-major axis and eccentricities.

A parallel programming based numerical code was written in Fortran with standard MPI protocols in order to solve the Equations (3-11) with double precision in an iterative manner to deduce the orbits of the planets and planetesimals. The orbital trajectories for the eight planets were deduced at a time-step of 8 hours. Based on these trajectories, an assessment was made for the acceleration vectors of the planetesimals. Rather than performing it for (N-10) number of planetesimals in a serial computational manner by a single microprocessor, the planetesimals were assigned to $M_p$ number of microprocessors with each processor executing computation for $(N-10)/M_p$ planetesimals. This resulted in faster computational speed. The detailed orbital trajectories of all the planetesimals and planets were followed in terms of perihelion and aphelion, semi-major axis, eccentricities during the simulation runs. Based on the deduced orbits of the planetesimals, the probability distribution function ($P^i_k$) for finding the planetesimals at distinct heliocentric distances, especially, within the four selected annular rings of interest were estimated. $P^i_k$ determines the probability of finding the planetesimals within a $k^{th}$ annular ring, starting initially from the $i^{th}$ annular ring. The probability of planetesimals collision, $P_{Coll.}$ and $f^i_k$ (the mass fraction of the collision debris contribution in the $k^{th}$ annular ring from the planetesimals that were initially introduced in the $i^{th}$ annular ring) are both proportions to $P^i_k$. Thus, in a manner, this probability estimate based on our simulations will determine the extent of homogenization among the distinct chemical reservoirs. The probability is estimated by generating the entire closed orbits of the planetesimals around 2π radians, at a specific time of interest from their semi-major axis and eccentricities. Further, an assessment is made regarding the most-likelihood of finding a specific planetesimal along its orbit at a given heliocentric distance. This process is repeated for all the planetesimals introduced initially among the four annular rings. Based on these individual probabilities, the most likelihood normalized probability (in percentage) is deciphered for any given planetesimal, introduced initially within a specific annular ring, to be present at any given heliocentric distance. The detailed normalized probability distribution functions are presented in Tables 2 and 3 for the two major simulation Sets A and C in the supplementary file. These individual probabilities add up to 100 % due to normalization, and the zero probabilities represent absence of the planetesimals in a specific region. Apart from the major four annular asteroid belt rings of our prime interest, we have also presented the probabilities for finding the planetesimals across the entire solar system in terms of orbital cross-overs with the planets and the interplanetary space. We cannot propagate error in numerical simulation in an *ab initio* manner on the basis of the error contributions due to various assumptions. This aspect can only be considered as a qualitative argument in the limitations of the various assumptions.

The gravitational perturbation of the planets experienced by the planetesimals, especially around the prominent Kirkwood gaps result in a gradual increase in eccentricities of the planetesimals. This leads to a gradual mixing of the two taxonomy classes across the four heliocentric annular rings. The orbital crossing of planetesimals with high eccentricities raises the probability of collisions among the planetesimals across distinct composition types. This increases the likelihood of the growth of large planetesimals from smaller planetesimals of distinct compositions at a macro-scale.

Finally, as mentioned earlier in order to consider the gravitational influence of planets on the orbital mixing of the planetesimals from distinct bulk chemical composition reservoirs in the asteroid belt we considered gradual accretion of planets (Table 1) over the initial 10 Myrs. in the case of simulation sets BI-II and CI-II according to a simplified Equation (12) (*Assumption # 12*).

$$M_P(t) = M_{P-Final}\left(1 - e^{\frac{-t}{t_{acc.}}}\right) \tag{12}$$

Here, $M_{P\text{-Final}}$ is the final mass of a planet, and $M_P(t)$ is the accreted mass at time 't' during accretion. The time, '$t_{acc.}$' is the characteristic time of accretion. It is assumed to be 2 Myrs. for Mercury and Mars, 4 Myrs. for the giant planets and 7 Myrs. for Earth and Venus on the basis of the estimates obtained for the planetary accretion timescales (e.g., Kruijer *et al.* 2017; Marchi *et al.* 2020; Schiller *et al.* 2020). We have not considered the migration of planets in the present work. The underlying assumption (*Assumption # 13*) associated with the accretion of planets definitely presents an oversimplified picture. However, the mathematical form of the equation is generally applicable in the case of physical processes that show rapid growth during their initial stages followed by reduction in the growth eventually leading to its saturation. The planets were formed as a result of runaway growth of planetary embryos. The growth rate substantially reduced at the final accretion stages during which the planets acquired their final masses. The gravitational influence of the mass that has not accreted on the planets at any given time was also not taken into account. Prior to its accretion, this mass can be assumed to have been uniformly distributed across the annular rings at planetary heliocentric distances, specifically, in the form of nebula gas in the case of the giant planets Jupiter and Saturn, such that its net resultant force on any test mass particle could be zero (*Assumptions # 14*). The gaseous giants, Jupiter and Saturn, exert the maximum gravitational influence on the planetesimals dynamics and most of their masses were in the form of nebula gases rich in hydrogen and helium. We cannot incorporate the aspects related with the interaction of the accreting giant planets with the nebula gas as it is beyond the scope our present work (*Limitation #14*). Even if the mass of the planets is in the form of planetary embryos that are not uniformly distributed, their net gravitational influence cannot be substantial compared to their final accreted state as a planet.

Among the 14 major assumptions considered in the work, all the assumptions except for the assumptions 2-4 can be relaxed by improving the computational resources. The $2^{nd}$ assumption assumes traditional model of asteroid belt formation, whereas, the $3^{rd}$ assumption requires development of theoretical formulation for understanding the dynamical processes related with the collision of planetesimals, followed by re-accretion of debris. The $4^{th}$ assumption is based on present observations of the asteroid belt. The majority of the earlier works in the field invariably made use of some of these assumptions in order to focus on the specific requirement of their simulation.

### 3. Results and Discussion

We have achieved the development of a N-body numerical simulation code for the gravitational interaction of planetesimals and asteroids due to the sun and the accreting planets. Further, we have tried to study the orbital dynamical evolution of the planetesimals from the two NC (S- type) and CC (C-type) reservoirs in the asteroid belt due to the gravitational influence of the accreting planets. Based on the orbital mixing of planetesimals across the four annular rings an assessment regarding the average chemical evolutionary trends can be made in a semi-quantitative manner based on the Equations (1, 2) . As the ensemble of planetesimals evolves within a specific ring due to collision induced disruptions and re-accretion, the finally accreted planetesimals/asteroids will theoretically acquire bulk composition representing the diverse planetesimal fragments. The results obtained for the set of simulations described in the Table 1 are presented in Figs. 1-6. These figures contain the temporal evolution of the orbital parameters across the asteroid belt. Each set of simulation (Table 1) was analyzed with two distinct initial chemical composition distributions among the four annular rings.

The dynamical orbital evolution of the planetesimals across the four annular rings at different time-steps of the three simulation sets is presented in Figs. 1, 3, 5. In the case of Set A simulation (Figure 1), the eccentricity excitation due to gravitational interaction of the giant planets (Jupiter and Saturn) commence within the initial 100 years. The simulation set

A constitutes our basic test run simulation that was used for the development of the code. It represents the contemporary solar system where the asteroids are introduced as test particles in the solar system with already formed planets. In this simulation, the planets are assumed to have instantaneously formed with their final masses and contemporary value of eccentricities at the beginning of this simulation. The eccentricity excitations become prominent within the initial 1 Myr. due to the adopted choice of the planets masses and eccentricities. The dynamical evolution almost achieves saturation in terms of the net planetesimal loss from the various annular rings at resonances well within the initial 5 Myrs. It acquires steady-state within the initial 4-5 Myr. The prominent Kirkwood gaps appear around 1 Myrs. These gaps almost stabilize within ~5 Myrs. The behavior of the dynamical evolution of the asteroids deduced from the present work is verified by the clear observations of the Kirkwood gaps at resonances that are identical to the previous works. The maximum loss takes place from the 1$^{st}$ ring due to $\nu_6$ secular resonance. Since, this ring is assumed to be initially populated with the S-type (NC) planetesimals, the lost S-type asteroids are scattered throughout the interplanetary space in a predominant manner. The Table 2 in the supplementary file presents a detailed account of the percentage likelihood ($P^i_k$) of finding a planetesimal at any particular heliocentric distance in the solar system over time as defined through the equations (1, 2). The likelihood of finding the S-type asteroids that were originally introduced within the 1$^{st}$ ring drops to ~19 % within the initial 5 Myrs. On the contrary, the likelihood of finding the C-type (CC) asteroids that were originally introduced within the 4$^{th}$ ring drops to ~36 % during this time. Thus, Saturn plays a predominant role in shaping the inner asteroid belt and redistributing the S-type asteroid throughout the solar system. The accretion timescale of Saturn along with its eccentricity in the early solar system will thus play an important role in distributing the S-type (NC) planetesimals. This aspect is further discussed later.

We ran an additional simulation by reducing the number of test particles (planetesimals) initially introduced at the beginning of the simulations by a factor of ~5 as compared to the Set A simulation. This simulation also inferred all the salient features in terms of resonances that are observed in the case of Set A simulation (Figure 1). We have avoided the presentation of this simulation in this work for this very reason. This simulation was performed to test the performance of the numerical approach with lesser number of test particles. Since, the dynamical orbital evolutionary studies consume a lot of computational time even with a parallel processing computational grid, it is extremely important to optimize the number of test particles. Based on the experience gathered from this simulation, we optimized the total number of test particles for the Set B and Set C simulations (Table 1).

Apart from the detailed account of the normalized probability distribution of finding a planetesimal that is moved from the initial location in the asteroid belt to any place in the solar system (Table 2), we present graphically the percentage likelihood ($P^i_k$) of finding a planetesimal initially moved from a specific annular ring to any other annular ring in the neighborhood (Figure 2). This covers the heliocentric distance between the orbits of Mars and Jupiter. The total interplanetary region presented in the figure represents the entire interplanetary space and the region beyond the present orbit of Neptune. A substantial number of planetesimals/asteroids are lost beyond the orbit of Neptune.

As inferred from the Figure 2 for the simulation A, a substantial number of S-type planetesimals/asteroids initially introduced from the 1st annular ring can spend considerable time in the 2nd – 4th annular rings, and beyond due to the $\nu_6$ secular resonance as mentioned above. As we proceed to the planetesimals that are initially introduced within the 2nd ring, the likelihood of getting them raised in terms of orbits to the outer annular rings is reduced. The trend continues in the outer annular rings. The outer 3rd and 4th annular rings that are initially populated with the C-type (CC) planetesimals allow movement of their planetesimals mostly to the neighboring ring. There is less than 1 % likelihood that the planetesimals that originate initially from the 3rd and 4th annular rings are ejected out of the solar system at least during the initial 5 Myrs. In general, the maximum change in the orbital evolutionary trends is observed during the initial 1 Myrs. The evolutionary trends acquire almost saturation within 5 Myrs. at least in the cases where the significant percentage likelihoods are ≥ 1%. As argued in the previous section, the probability of planetesimals collision, $P_{Coll.}$ and $f^i_k$ (the mass fraction of the collision debris) as defined by equations (1, 2) are both proportional to the deduce $P^i_k$. These parameters in turn determine chemical homogenization. Thus, the Figure 2 provides a basis for determining the likelihood of generating planetesimals of varied compositions across the end-member NC-CC (S- and C-type) dichotomy mixing line at least in a semi-quantitative manner. The deduced normalized probabilities (Figure 2; Table 2) provide an estimate regarding the chemical nature of planetesimals that could have formed as a result of collision induced fragmentation and re-accretion of the two diverse constituents, NC and CC, in case the system achieved complete equilibrium in terms of collisions, fragmentations and re-accretion of planetesimals. In Set AI simulation, with the 1st ring and 2nd-4th rings having an initial NC and CC compositions, respectively, the percentage contribution ratios of NC and CC planetesimals in the 1st annular ring will evolve from ~67:21 % at 0.1 kyr. (kiloyear) to ~19:19 % at 5 Myr (Figure 2). This could result in a major mixing among the two dichotomy chemical reservoirs, thereby, almost obliterating the initially creating NC reservoir in the ring. The 2nd ring will evolve from 15:87 % at 0.1 kyr. to 7:50 % contributions of NC and CC planetesimals. This includes the possible likelihood contributions from the inner and outer rings. The percentage probabilities are based on the initial number of planetesimals of a specific type introduced within the four individual rings at the beginning of the simulations. These initial numbers are assumed to be same in all the annular rings. The 3rd and the 4th rings will mostly maintain the CC dominance in the Set AI simulation as the CC planetesimals from the neighboring rings can alone orbit through these rings.

In Set AII simulation, the 1st ring will maintain a majority in the NC planetesimals as the 2nd ring that mostly contributes to 1st ring happens to be initially populated by NC planetesimals. In terms of the NC to CC planetesimals ratio, the 2nd ring in the case of Set AII simulation will evolve from ~75:25 % at 0.1 kyr. (kiloyear) to 56:29 % at 5 Myr. The 3rd ring, initially populated by CC planetesimals, will evolve from ~18:80 % at 0.1 kyr. (kiloyear) to 15:74 % at 5 Myr. The 4th ring remains CC dominated in this simulation. Based on these analyses, we anticipate a substantial mixing among the NC (S-type) and CC (C-type) reservoirs if we invoke spontaneous accretion of giant planets with their contemporary eccentricities. The accretion of diverse planetesimal populations with these homogenized compositions prior to 1.5 Myrs. could have produced wide-scale melting and planetary differentiation in case $^{26}$Al was present in these various reservoirs. This could have possibly obliterated the observed dichotomy of the NC and CC achondrites and iron-parent body meteorites (Kleine *et al.* 2020; Schneider *et al.* 2020; Burkhardt *et al.* 2021).

In Set B simulation, the planets acquired mass gradually according to Equation (12) that is based on the deduced accretion rates of the planets. The accretion of these planets commence with the rapid formation of the rocky cores of the planets with masses 5-10 times larger than Earth. The subsequent accretion of the nebular hydrogen and helium occurs gradually over the prescribed timescales. In context to the formation of the solar system the scenario Set B provides the most realistic scenario as far as the planet formation is concerned. The initial eccentricities of the planets are assumed to be zero in this case. There is a very limited observed excitation in the eccentricities of the planetesimals (Figure 3) during the initial 10 Myrs. The excitation is mostly confined to the 2:1 and 3:1 orbital resonances with Jupiter. An extremely small magnitude of $\nu_6$ secular resonance is observed. Since, the majority of the planetesimals are mostly confined to their original orbits (Figure 3), an insignificant change in the temporal evolution of the extent of planetesimals crossing across the various annular rings is observed (Figure 4). This simulation in fact verifies the robustness of our numerical technique. Due to the $\nu_6$ secular resonance an increase in the net contribution of the S-type (NC) planetesimals ejected out of the solar system from the 1$^{st}$ annular ring is observed over time.

In order to excite the eccentricities of the planetesimals it is essential to excite the eccentricities of the planets undergoing accretion growth. We ran a Set C simulation with the assumed contemporary values of the eccentricities of planets. The planets accreted mass gradually on the basis of Equation (12). The giant planets would have acquired ~22 % and ~53 % of their final masses within the initial 1 and 3 Myrs., respectively. As a result, a significant excitation in the eccentricities of the planetesimals is observed within the initial 1 Myrs. (Figure 5). The gradual increase in the eccentricity takes place around the $\nu_6$ secular resonance, 3:1, 5:2, 7:3 and 2:1 resonances over the initial 10 Myr., the time duration over which the presence of $^{26}$Al determines whether a planetesimal will undergo differentiation, thermal metamorphism or aqueous alteration or simply survives as rubble pile assemblage. The orbital crossing of the planetesimals across the distinct annular rings is substantially large (Figure 6) compared to Set B simulation. However, in the early stages, the eccentricity excitations are substantially small compared to those observed in the Set A simulation (Figure 1). The $\nu_6$ secular resonance is specifically small during the initial ~2 Myrs. due to the low mass of the accreted Saturn. The Table 3 in the supplementary file presents the normalized percentage likelihood ($P^i_k$) of finding a planetesimal at any particular heliocentric location in the solar system over time. In Set CI simulation (Figure 6 and Table 3), the percentage contribution ratios of NC and CC planetesimals in the 1$^{st}$ annular ring will evolve from ~67:21 % at 0.1 kyr. (kiloyear) to ~50:21 % at 5 Myr and further to ~37:19% at 10 Myr. The 2$^{nd}$ ring in Set CI will evolve from 15:87 % at 0.1 kyr. to 12:85 % at 5 Myrs. and further to ~9:83% at 10 Myr in terms of the contributions of NC and CC planetesimals. Compared to the simulation Set AI, this is a substantial improvement in terms of maintaining the observed dichotomy (Kleine *et al.* 2020; Schneider *et al.* 2020; Burkhardt *et al.* 2021), specifically, during the initial 5 Myrs. during which NC and CC achondrites and chondrites accrete and evolved thermally. The 3$^{rd}$ and the 4$^{th}$ rings mostly maintain the CC dominance. Further, in the case of Set CII simulation, the 2$^{nd}$ ring evolves from ~75:25 % at 0.1 kyr. (kiloyear) to 67:29 % at 5 Myr. and further to ~62:30% at 10 Myr. in terms of contributions of NC and CC planetesimals. The 3$^{rd}$ ring will evolve from ~17:80 % at 0.1 kyr. (kiloyear) to 17:76 % at 5 Myr. and further to ~15:76 % at 10 Myr. The 4$^{th}$ ring remains CC dominating.

On the basis of the three distinct set of simulations A, B and C, it can be deduced that in order to circumvent substantial orbital cross-mixing of planetesimals in the early solar system across distinct heliocentric annular rings, the accretion growth of the giant planets should be prolonged. Further, this should be supported by low eccentricity values of the giant planets as compared to their contemporary values. The prolonged accretion timescale of giant planet with zero eccentricities restricts the substantial orbital crossover mixing of planetesimals of distinct compositions, especially, during the initial couple of million years of the solar system. In general, the observed chemical dichotomy can be preserved in case the giant planets grew mass gradually with initially low eccentricity values, or else, the direct gravitational collapse due to streaming instability of turbulent pebble clouds rather than the accretion of small planetesimals resulted in the formation of the majority of the asteroids (Klahr & Schreiber 2020) from localized pockets of different compositions. The migration of giant planets could complicate this deduction in case the accreting giants Jupiter and Saturn had been closer to the asteroid belt during the initial stages of couple of million years. This could have caused substantial inter-ring mixing of the planetesimals of different compositions.

In the above analyses, we consider the possibility of the average evolving trends prevailing within a ring in terms of the compositions of the S-type and C-type asteroids/planetesimals in a semi-quantitative manner. However, these hypothesized averaged-out compositions are based on the bulk compositions of planetesimals in case the system achieved complete equilibrium in terms of collisions, fragmentations and re-accretion of planetesimals. However, in case the system is far away from such assumed equilibrium, it is quite possible that within an annular ring at a specific time, an array of compositions is generated by collision induced growth of planetesimals with distinct proportions of S- (NC) and C-type (CC) planetesimals. This would produce an evolutionary pattern in the form of bands rather than averaged out evolutionary trends. Further, the collision induced debris from the planetesimals could result in planetesimal fragments with high eccentricities. These fragments could result in large orbital crossovers of the planetesimals of distinct compositions across the different annular rings. This could result is larger chemical homogenization across annular rings with distinct initially assumed compositions. Further, since the present work broadly demarcates the NC and CC dichotomy types, the meteoritic sub-classification, e.g., the distinct carbonaceous (CV, CM, CO, CR, CH, etc,), ordinary (H, L, LL) and enstatite (EL, EH) chondrites cannot be examined within the present preview. There is a possibility that certain localized regions were responsible for the formation of these distinct reservoirs with wide-ranging chemical and isotopic compositions. The streaming instabilities among localized turbulent pebble clouds (Klahr & Schreiber 2020) could have directly led to the accretion of asteroids within these reservoirs.

Our simulations are based on the traditional model of the *in situ* growth of asteroids within the asteroid belt (*Assumption # 2*). However, in the case of the alternative model, the Grand Tack model, the initial process of populating the asteroid belt from the inner and the outer regions of the solar system by the migrating giant planets, the various annular rings, specified in the present work, could have resulted in a population admixture of S- and C- type planetesimals. The inner region could have acquired more S-type along with C-type planetesimals compared to the outer region accumulating mostly C-type planetesimals. This model could have resulted in a chemical mixing for the chondrites, achondites and iron meteorite parent bodies across the NC-CC dichotomy end-members. The details regarding the extent of the mixing have to be independently worked out on the basis of detailed simulations for the concerned scenario. However, in case the processes associated with the asteroid feeding to the belt could have continued for several million years (~2-5 Myrs.) during which the ordinary, enstatite and carbonaceous chondrites, etc., were predominately feed into the asteroid belt from their specific region of origin, the extent of mixing across the end-members would have been limited, specially, in the case of achondrite and iron meteorite parent bodies from the NC and CC achondrites that experienced differentiation within the initial 1.5 Myr.

Finally, it is important to understand the timeframe of collision induced homogenization of planetesimals of distinct compositions across the distinct annular rings in context to the role of $^{26}$Al as the main source of heat for planetary differentiation, thermal metamorphism, aqueous alteration and thermal sintering of the planetesimals. The planetesimals formed from these evolving ensembles within distinct annular rings may experience planetary scale differentiation, exclusively due to $^{26}$Al, within the initial ~1.5 Myrs (Sahijpal et al., 2007). We are assuming an insignificant heat contribution from $^{60}$Fe due to its low abundance in the early solar system. The mixed compositional reservoirs encompassed within these bodies will thus provide the average composition for the achondrites, iron and stony-iron meteorite parent bodies. In case the accretion of planetesimals occurred beyond the time-span of ~1.5 Myrs., it could have resulted in thermal metamorphism or aqueous alteration. The accretion of planetesimals occurring beyond ~5 Myrs. (Sahijpal, 2022) could have produced unconsolidated rubble-pile asteroids with low density (Walsh, 2018). It should be noted that these processes could have occurred along with the nebular processes dealing with the formation of chondrules during the initial ~2-4 Myrs. This could have occurred within the same region.

4.  **Summary**

Numerical simulations have been performed for the dynamical evolution of the planetesimals among the NC (S-type) and CC (C-type) chemical reservoirs in the early solar system in order to understand the origin and evolution of chondrites, achondrites and iron meteorite parent bodies from these two distinct broad reservoirs. We numerically considered the initial formation of NC and CC chemical reservoirs within the asteroid belt in the form of annular rings, with the former concentrated within the inner belt. N-body numerical simulations of the dynamical orbital evolution of the S-type (CC) and C-type (CC) planetesimals have been performed. An ensemble of planetesimals was dynamically evolved under the gravitational influence of the sun and the accreting planets. The objective is to understand the possibility of the extent of homogenization of distinct chemical compositions from the assumed existing two end-member reservoirs during the initial 10 Myrs in a semi-quantitative manner. This would numerically constrain the conditions that would preserve the observed dichotomy. Our simulations impose constraints on the rate of accretion growth of the planets, Jupiter and Saturn along with their initial eccentricities in the early solar system. The observed dichotomy among the NC and CC chondritic and achondritic reservoirs is maintained at least during the initial ~5 Myrs. provided the accretions of Jupiter and Saturn occurred at least over an identical timescale, with probably lower values of the two planets eccentricities compared to their contemporary values.

**Acknowledgement:** We are extremely grateful to the various comments received from the reviewer that led to the major improvement of the manuscript. We are extremely grateful to the IUCAA, Pune, computational network grid where the parallel computational processing of the present work was performed.

**Appendix A.** Supplementary data to this article contains Tables 2 and 3 corresponding to the simulation sets A and C.

**Table 1.** Simulation details of the various dynamical models.

| S.No. | Model[a] | Initial number of particles per ring | Simulation nature in terms of major planets |
|---|---|---|---|
| 1 | Set (AI, AII) | 2908 | Planets initially introduced with final masses, present eccentricities and semi-major axes |
| 2 | | | |
| 3 | Set (BI, BII) | 1088 | Planets introduced with gradual increasing masses and zero eccentricities. |
| 4 | Set (CI, CII) | 1088 | Planets introduced with gradual increasing masses and present eccentricities. |

[a] The asteroid belt is divided into four annular rings, viz., 1st ring (1.8-2.2 AU), 2nd ring (2.2-2.6 AU), 3rd ring (2.6-3.0 AU) and 4th ring (3.0-3.4 AU). Each ring contains equal number of test particles in the beginning. In the Set AI-CI, the inner 1st ring and the outer 2nd - 4th annular rings were assumed to be initially populated with S-type (NC) and C-type (CC) planetesimals, respectively. In the Set AII-CII, the inner 1st - 2nd annular rings and the outer 3rd – 4th annular rings were assumed to be initially populated with S-type (NC) and C-type (CC) planetesimals, respectively.

**Figures**

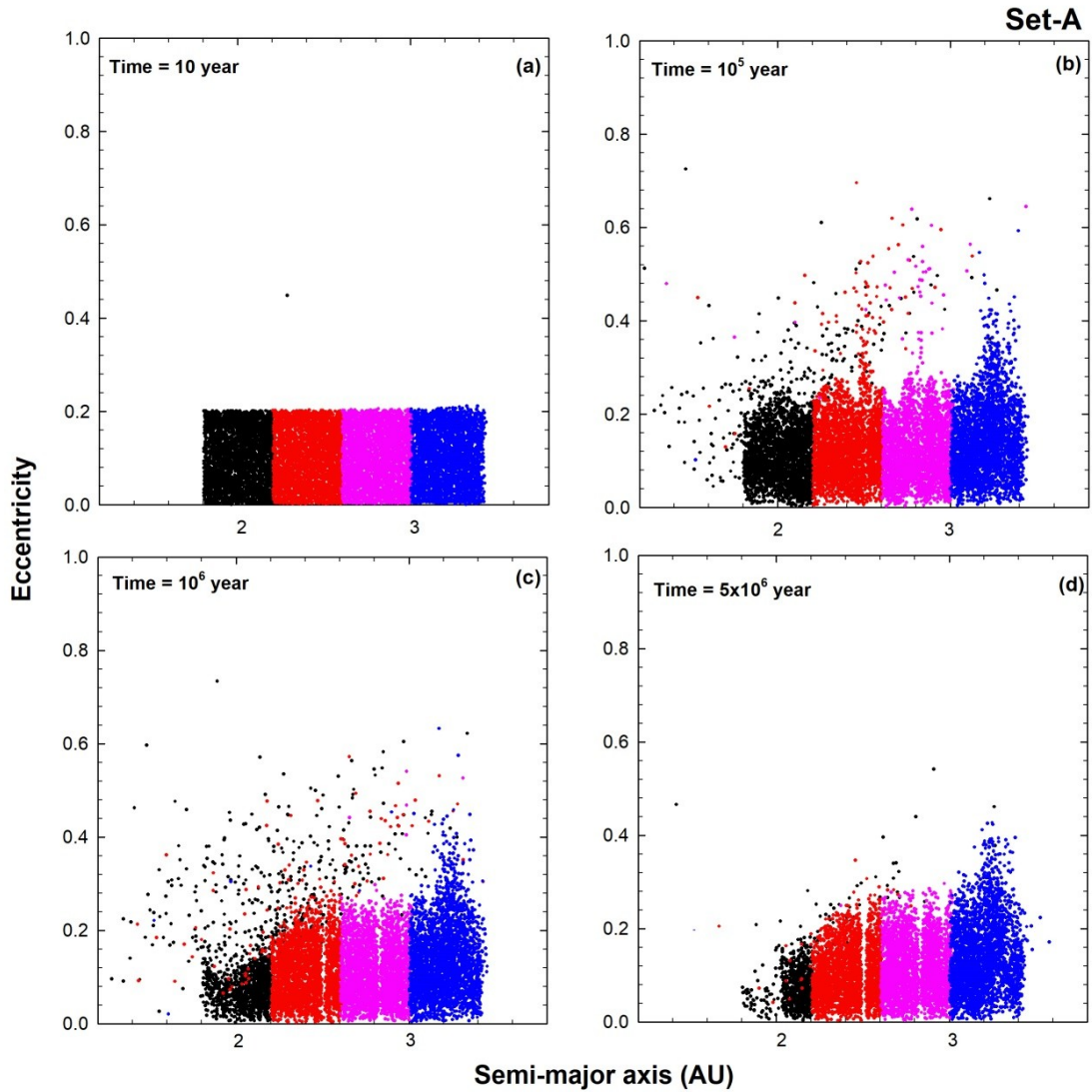

**Figure 1.** The dynamical orbital evolution of planetesimals (asteroids) in terms of the semi-major axis (AU) and eccentricities over 5 Myrs. for the Set A simulations. In the Set AI simulation, the black color dots represent planetesimals with S-type composition and the red, magenta and blue color dots represent planetesimals with C-type composition. In the Set AII simulation, the black and red color dots represent planetesimals with S-type composition and the magenta and blue color dots represent planetesimals with C-type composition.

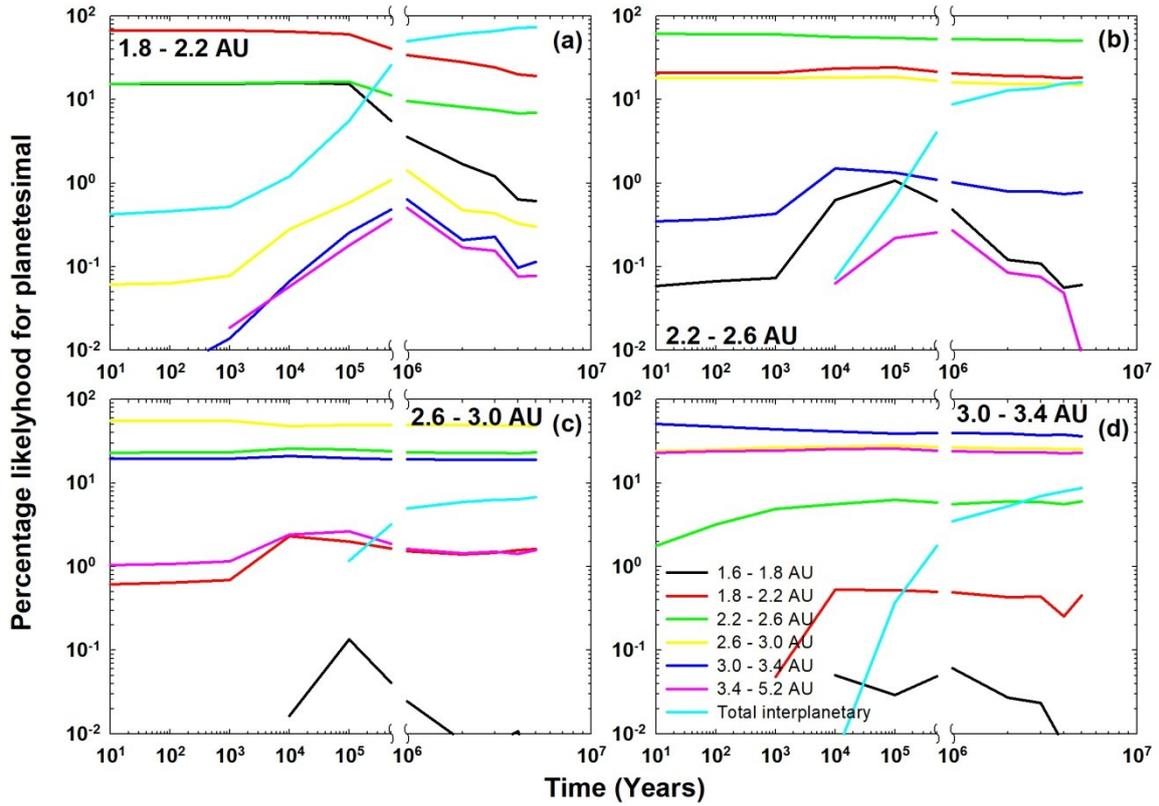

**Figure 2.** The temporal evolution of the initially populated planetesimals from the four annular rings, a) 1st ring (1.8-2.2 AU), b) 2nd ring (2.2-2.6 AU), c) 3rd ring (2.6-3.0 AU) and d) 4th ring (3.0-3.4 AU) in case of Set A simulation. The evolution is presented in terms of the percentage likelihood ($P^i_k$) that a planetesimal (asteroid) with the initial semi-major axis from a specific ring, 'i', will crossover to a specific heliocentric region, 'k'. The total interplanetary region represents the entire interplanetary space and the region beyond Neptune.

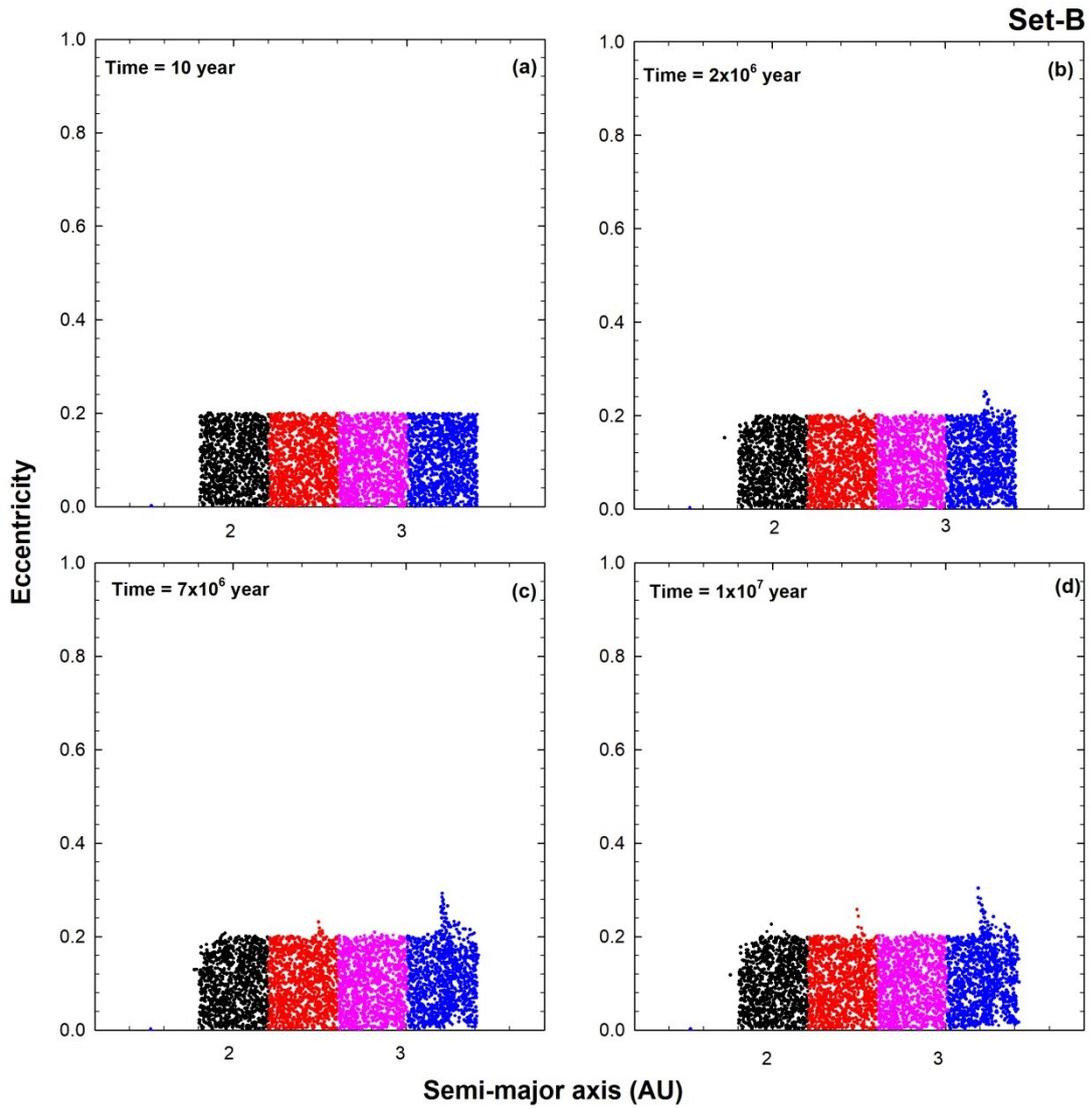

**Figure 3.** The dynamical orbital evolution of planetesimals (asteroids) in terms of the semi-major axis (AU) and eccentricities over 10 Myrs. for the Set B simulations. In the Set BI simulation, the black color dots represent planetesimals with S-type composition and the red, magenta and blue color dots represent planetesimals with C-type composition. In the Set BII simulation, the black and red color dots represent planetesimals with S-type composition and the magenta and blue color dots represent planetesimals with C-type composition.

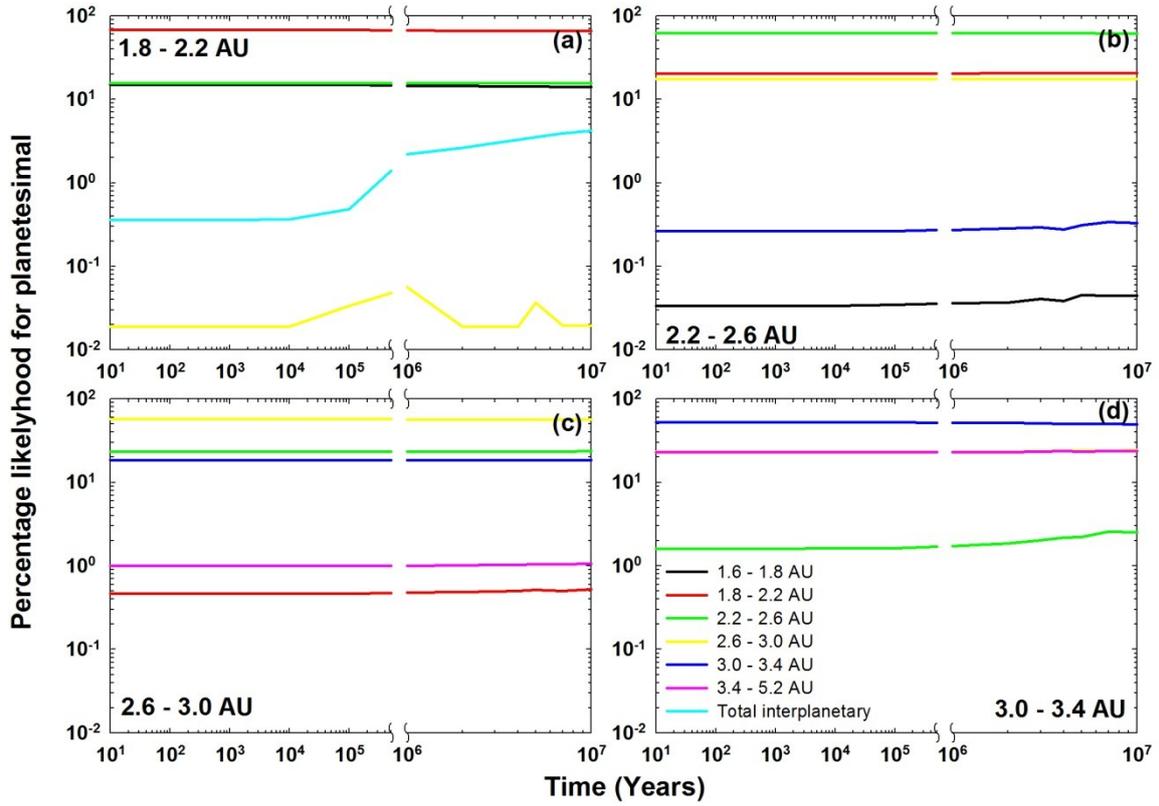

**Figure 4.** The temporal evolution of the initially populated planetesimals from the four annular rings, a) 1st ring (1.8-2.2 AU), b) 2nd ring (2.2-2.6 AU), c) 3rd ring (2.6-3.0 AU) and d) 4th ring (3.0-3.4 AU) in case of Set B simulation. The evolution is presented in terms of the percentage likelihood ($P^i_k$) that a planetesimal (asteroid) with the initial semi-major axis from a specific ring, 'i', will crossover to a specific heliocentric region, 'k'. The total interplanetary region represents the entire interplanetary space and the region beyond Neptune.

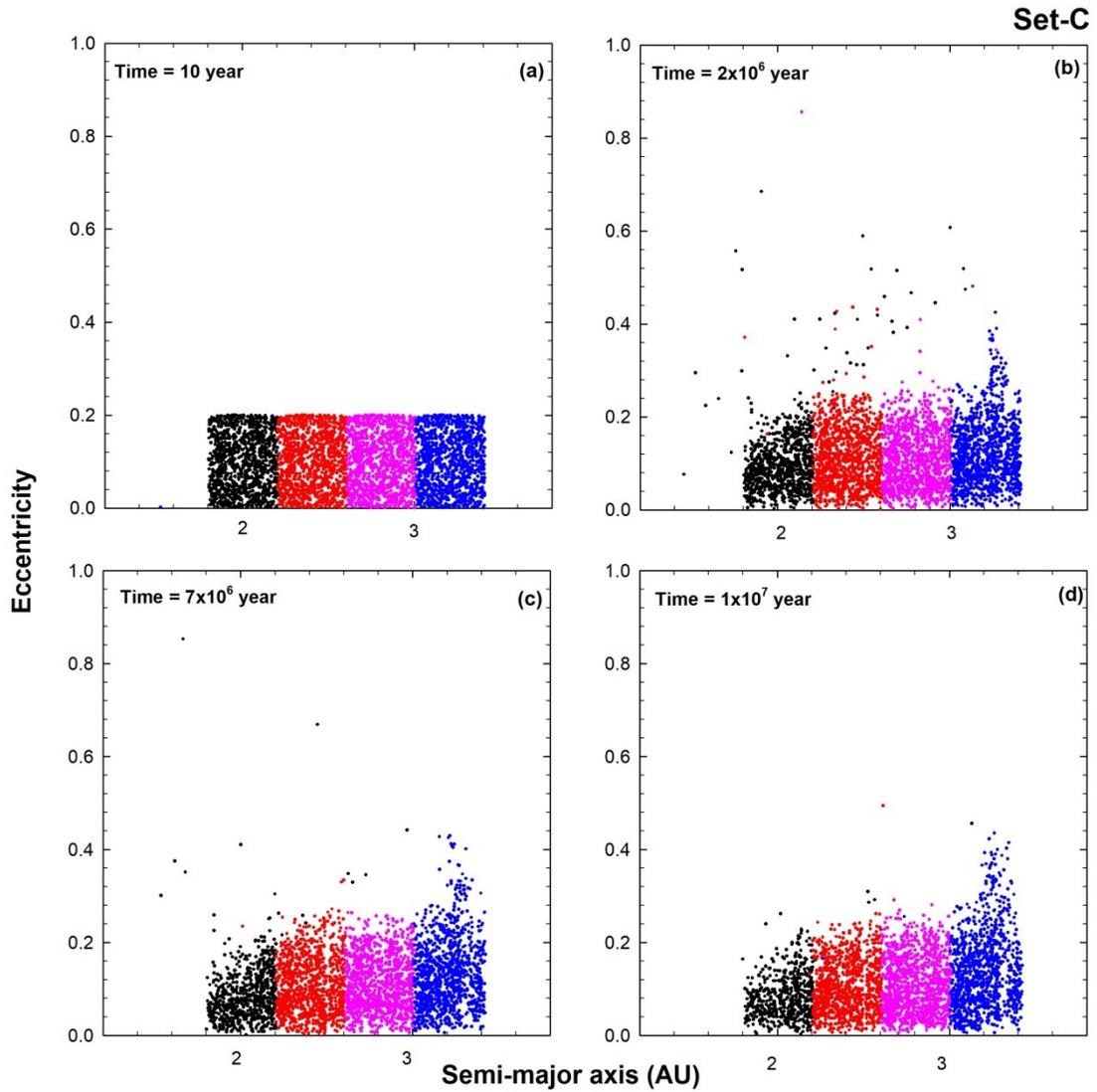

**Figure 5.** The dynamical orbital evolution of planetesimals (asteroids) in terms of the semi-major axis (AU) and eccentricities over 10 Myrs. for the Set C simulations. In the Set CI simulation, the black color dots represent planetesimals with S-type composition and the red, magenta and blue color dots represent planetesimals with C-type composition. In the Set CII simulation, the black and red color dots represent planetesimals with S-type composition and the magenta and blue color dots represent planetesimals with C-type composition.

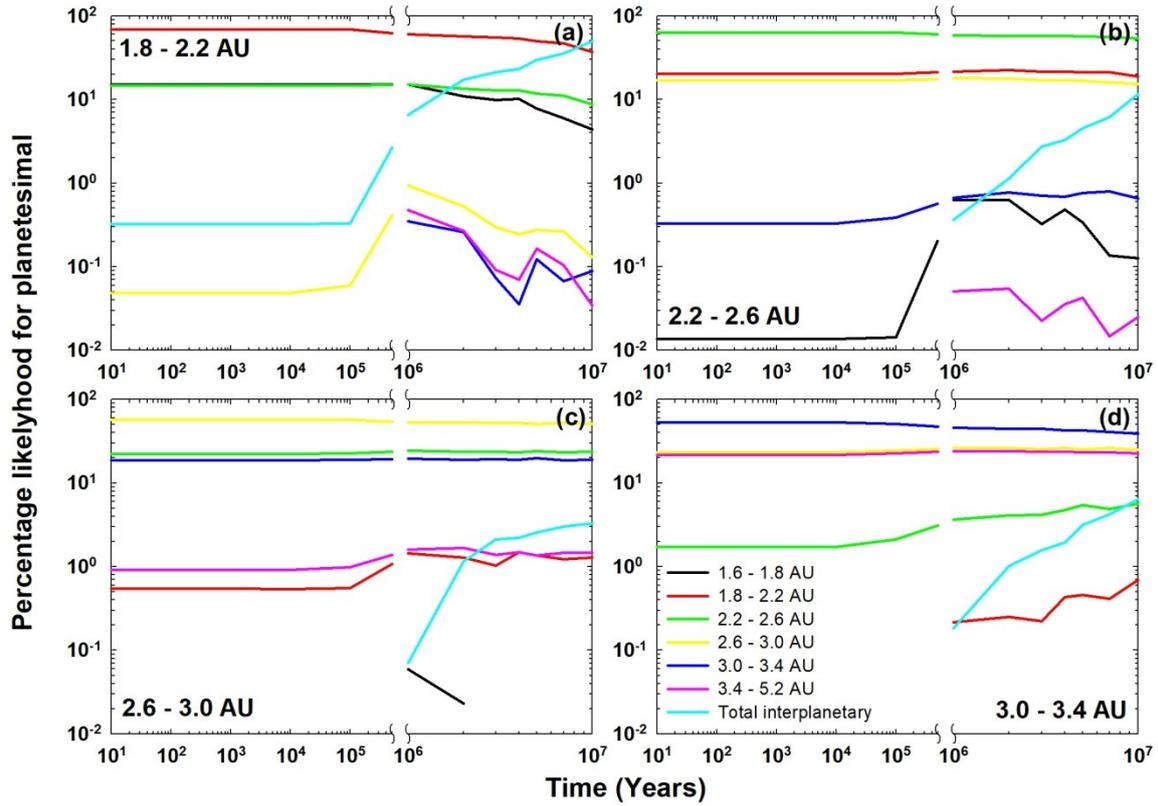

**Figure 6.** The temporal evolution of the initially populated planetesimals from the four annular rings, a) 1$^{st}$ ring (1.8-2.2 AU), b) 2$^{nd}$ ring (2.2-2.6 AU), c) 3$^{rd}$ ring (2.6-3.0 AU) and d) 4$^{th}$ ring (3.0-3.4 AU) in case of Set C simulation. The evolution is presented in terms of the percentage likelihood ($P^i_k$) that a planetesimal (asteroid) with the initial semi-major axis from a specific ring, 'i', will crossover to a specific heliocentric region, 'k'. The total interplanetary region represents the entire interplanetary space and the region beyond Neptune.